\documentclass[aps,prl,twocolumn,preprintnumbers,showpacs,amsmath,amssymb,superscriptaddress]{revtex4}
\usepackage{graphicx}
\usepackage{dcolumn}
\usepackage{bm}

\begin{document}


\title{Experimental Synchronization of Independent Entangled Photon Sources}

\author{Tao Yang}
\thanks{These two authors contribute equally to this work.}
\affiliation{%
Department of Modern Physics and Hefei National Laboratory for Physical Sciences
at Microscale, University of Science and Technology of China, Hefei, Anhui 230026, China%
}
\author{Qiang Zhang}
\affiliation{%
Department of Modern Physics and Hefei National Laboratory for
Physical Sciences
at Microscale, University of Science and Technology of China, Hefei, Anhui 230026, China%
}
\author{Teng-Yun Chen}
\affiliation{%
Department of Modern Physics and Hefei National Laboratory for
Physical Sciences
at Microscale, University of Science and Technology of China, Hefei, Anhui 230026, China%
}
\author{Shan Lu}
\affiliation{%
Department of Modern Physics and Hefei National Laboratory for
Physical Sciences
at Microscale, University of Science and Technology of China, Hefei, Anhui 230026, China%
}
\author{Juan Yin}
\affiliation{%
Department of Modern Physics and Hefei National Laboratory for
Physical Sciences
at Microscale, University of Science and Technology of China, Hefei, Anhui 230026, China%
}\author{Jian-Wei Pan}
\thanks{Email:pan@ustc.edu.cn}
\affiliation{%
Department of Modern Physics and Hefei National Laboratory for
Physical Sciences
at Microscale, University of Science and Technology of China, Hefei, Anhui 230026, China%
}
\author{Zhi-Yi Wei}
\thanks{These two authors contribute equally to this work.}
\affiliation{%
Laboratory of Optical Physics, Institute of Physics, Chinese Academy of Sciences,
Beijing 100080, China%
}
\author{Jing-Rong Tian}
\affiliation{%
Laboratory of Optical Physics, Institute of Physics, Chinese
Academy of Sciences,
Beijing 100080, China%
}
\author{Jie Zhang}
\affiliation{%
Laboratory of Optical Physics, Institute of Physics, Chinese
Academy of Sciences,
Beijing 100080, China%
}

\date{\today}

\begin{abstract}
Realistic linear quantum information processing necessitates the
ability to synchronously generate entangled photon pairs either at
the same or at distant locations. Here, we report the experimental
realization of synchronized generation of independent entangled
photon pairs. The quality of synchronization is confirmed by
observing a violation of Bell's inequality with 3.2 standard
deviations in an entanglement swapping experiment. The techniques
developed in our experiment will be of great importance for future
linear optical realization of quantum repeaters and quantum
computation.
\end{abstract}

\pacs{03.65.Ud, 42.65.Lm, 42.62.Eh}
\maketitle

Entangled photon pairs are an essential resource for linear optics
quantum information processing (LOQIP)
\cite{knill:40946,pan:1067}. For example, using linear optical
elements one can combine entanglement swapping
\cite{zukowski:4287} and entanglement purification
\cite{bennett:722} to efficiently generate highly entangled states
between two distant locations \cite{briegel:5932}. Moreover, one
can exploit linear optics and entangled photon pairs to achieve
logic operations between single photons \cite{pittman:032316}. On
this basis, one can further prepare cluster states to perform
one-way quantum computation
\cite{raussendorf:5188,nielsen:040503}. Recently, using entangled
photon pairs created by one and the same laser pulse significant
progress has been made in proof-in-principle demonstration of
entanglement swapping \cite{pan:3891}, entanglement purification
\cite{pan:417} and photonic logic operation
\cite{sanaka:017902,gasparoni:020504,zhao:030501}. However, in
reality scalable LOQIP necessitates the ability to synchronously
generate entangled photon pairs either at the same or at distant
locations \cite{knill:40946,pan:1067}. Here, we report
experimental realization of synchronized generation of independent
entangled photon pairs. The quality of synchronization is
confirmed by observing a violation of Bell's inequality with 3.2
standard deviations in an entanglement swapping experiment.

Entanglement swapping, i.e. teleportation of entanglement
\cite{zukowski:4287,bennett:1895}, is a way to project the state
of two particles onto an entangled state while no direct
interaction between the two particles is required. During
entanglement swapping, if each of the two particles is originally
entangled with one other partner particle, a Bell-state
measurement of the partner particles would thus collapse the state
of the two particles into an entangled state, even though they are
far apart.

One important application of entanglement swapping, probably also
the most important application, is in long-distance quantum
communication \cite{gisin:145}. Due to the absorption and
decoherence of quantum channel, the cost for communication between
two distant parties increases exponentially with the channel
length. One excellent solution is to connect distant communicating
parties with quantum repeaters \cite{briegel:5932}: firstly
dividing the whole quantum channel into several segments, and then
performing entanglement swapping and entanglement purification.
Therefore, in realistic realization of quantum repeaters one has
to achieved entanglement swapping with synchronized entangled
photon sources among all distributed segments.

Nowadays, entangled photon pairs are usually created via
parametric down-conversion from a UV laser pulse. In this case,
the UV laser pulses in each distributed segment must be
synchronized. One natural solution is to split a single UV laser
pulse into N beams and then distribute them to each segment
\cite{Riedmatten:0409093}. However, such a naive solution is not a
scalable scheme. This is because the maximal output power of a
single laser is technically limited and the efficiency of the
scheme will thus exponentially decrease with the number of
segments. A practical solution is to utilize synchronized pump
lasers to prepare entangled pairs in each segment. Thereafter we
connect these pairs via entanglement swapping. Here, we are going
to report the first experimental realization of this kind, i.e.
entanglement swapping with independent entangled photon pairs that
are created by two synchronized femto-second lasers.

Considering two independent EPR sources, each emitting a pair of
polarization entangled photons synchronously. The expected state
of the system consisting of two independent pairs can be written
as:
\begin{equation}\label{eq:total1}
|\Psi\rangle_{total}=\frac{1}{2}(|H\rangle_{1}|V\rangle_{2}-|V\rangle_{1}|H\rangle_{2})
\otimes(|H\rangle_{3}|V\rangle_{4}-|V\rangle_{3}|H\rangle_{4}).
\end{equation}
Here photons 1 and 2 (3 and 4) are entangled in the anti-symmetric
polarization state $|\Psi^{-}\rangle$. Note that, hereafter we
exactly follow the notations as used in ref. \cite{pan:3891}. From
Eq. (\ref{eq:total1}), one can easily see there is no any
entanglement of any of photon 1or 2 with any of the photon 3 and
4.

Rearranging the terms by expressing photon 2 and photon 3 in the
basis of Bell state, Eq. (\ref{eq:total1}) can be expressed as
\begin{eqnarray}\label{eq:total2}
|\Psi\rangle_{total} & = &
\frac{1}{2}(|\Psi^{+}\rangle_{14}|\Psi^{+}\rangle_{23}
+|\Psi^{-}\rangle_{14}|\Psi^{-}\rangle_{23}
\nonumber \\
& & +|\Phi^{+}\rangle_{14}|\Phi^{+}\rangle_{23}
+|\Phi^{-}\rangle_{14}|\Phi^{-}\rangle_{23}).
\end{eqnarray}
Eq. (\ref{eq:total2}) implies that projecting photons 2 and 3 in
one of the four Bell-states will lead the remaining photons 1 and
4 entangled in the corresponding Bell-state, despite they are
produced separately and never interacted with one another. Due to
the limitation of the linear optics element, only $50\%$
Bell-state can be analyzed. In our experiment we decide to analyze
only the case that photons 2 and 3 are projected in
$|\Psi^{-}\rangle_{23}$ state and interfering photons 2 and 3 at a
$50:50$ beam splitter is able to identify the
$|\Psi^{-}\rangle_{23}$ state. When detecting a coincident count
between the two detectors at the output ports of the beam
splitter, photons 2 and 3 are projected to $|\Psi^{-}\rangle_{23}$
state, and then photons 1 and 4 will be in the entangled state
$|\Psi^{-}\rangle_{14}$ .

Note that, since the Bell-state analysis relies on the
interference of photons 2 and 3 one has to guarantee the photons 2
and 3 have good spatial and temporal overlap at the beam splitter.
In previous experiments where the two photon pairs are created by
parametric down-conversion from the same laser pulse, the
interference of photons is guaranteed by making the coherence
times of interfering photons much longer than the pump pulse
duration \cite{Zukowski:75591}. However, since in our experiment
the two photon pair are created by parametric down-conversion from
two independent pump lasers, besides increasing the coherence
times of the interfering photons by inserting narrow bandwidth
filters in front of the detectors registering photons 2 and 3, one
has to further ensure that the two independent laser pulses are
synchronized perfectly and the timing jitter of synchronization is
much smaller than the coherence times. This is experimentally very
challenging.

Usually, femtosecond laser uses either active synchronization with
an electrical feedback device \cite{ma:021802}, or passive
synchronization by nonlinear coupling mechanism
\cite{Barros:18631}. In our experiment, we implement passive
technique to synchronize two Ti: sapphire lasers, because the
passive technique is stimulated by cross-phase modulation and
should be capable of operating at lower fluctuation, this  will
result in a very small timing jitter \cite{Wei:74171}. Considering
the two lasers operating at repetition frequencies of $f_1$ and
$f_2$ respectively before synchronized, they cross a Kerr medium
at repetition rate of $|f_1-f_2|$, and suffer frequency shift
according their temporal overlap in the Kerr medium, for example,
the slower pulse shifts to blue, and the faster one shifts to red.
Considering if the pulses start to cross inside the Kerr medium,
due to the negative group dispersions in the laser cavities, the
crossing of the two pulses will be enhanced after they take one
round trip in their cavities, when both cavities being adjusted to
be nearly equal. Therefore, the leading pulse will be slowed down
and the sluggish one will be fastened, until they overlap
maximally in time domain.

\begin{figure}[ptb]
\includegraphics[width=\columnwidth]{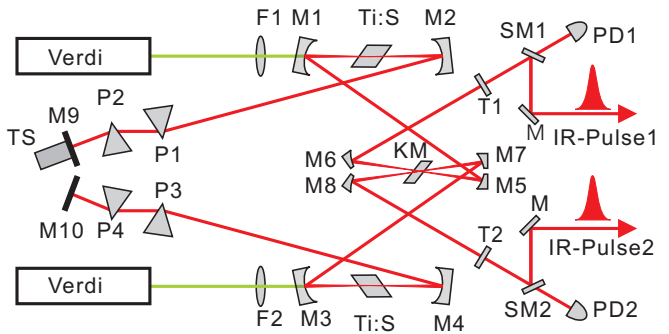}
\caption{Experimental setup of synchronized femtosecond pulse
lasers. F1 and F2 are lens to focus the pumping 532nm laser from
two Verdi laser systems; Ti1 and Ti2 are Ti:sapphire crystals; M1
- M10 are high reflection mirrors; P1-P4 are prisms; T1 and T2 are
output couplers with transmissivities of 20\%; On the top of the
figure, Ti1, M1-M5, P1, P2 and T1 constitute the first mode-lock
femtosecond laser cavity. An analogous mode-lock femtosecond laser
cavity shown at the bottom of the drawing is constituted of Ti2,
M6-M10, P3, P4 and T2. The two laser pulses are synchronized by
coupling both lasers in the Ti: Sapphire crystal Kerr medium (KM).
In order to induce stronger cross-phase modulation effect for
synchronization, we focus the beams in the Kerr medium and make
the two beams cross in the Kerr medium with a narrow angle.
Considering the crucial condition of synchronizing lasers, One end
mirror M5 is driven by a translation stage to match the two laser
cavity lengths. Both 788nm Infrared laser pulses are detected by
fast photodiodes (PD1 and PD2) behind beam samplers (BS1 and BS2).
Hence we can monitor the synchronization between two laser pulses
on an oscilloscope.} \label{fig1}
\end{figure}

In the experiment we synchronize two Ti:sapphire femtosecond
lasers by coupling both laser pulses in an additional Ti:sapphire
crystal. Figure1 is the schematic of the experimental setup of
laser synchronization. It consists of two Ti:sapphire femtosecond
lasers located at the top and bottom corners in Fig.\ref{fig1}
respectively. The symmetry of two cavities ensures that both
cavities length are the same, and both laser pulses work at the
same repetition rate of 81MHz, which provides the basic condition
of synchronization. To fine tune the match of cavities, a
translation stage (TS) is also used to drive the end mirror M5 of
the first Ti:sapphire laser. Both laser pulses are coupled into a
Ti:sapphire crystal KM to synchronize with each other. To enhance
the cross-phase modulation, focus mirrors M3 and M4 are inserted
into the first laser cavity, and M8, M9 are inserted into the
second laser cavity for introducing additional focal point inside
the KM.

We pump each Ti:sapphire femtosecond laser with a solid-state
diode-pumped 532-nm laser (Verdi-V10). Under the pump power of 8W
for each, each Ti:sapphire femtosecond provides 700mW power at
synchronized mode locked status, and the central wave lengths of
the lasers are 788nm. Thereafter, we measure the pulse durations
by auto correlator. The laser pulse durations (FWHM) are 60 fs and
70fs respectively. Further more, we measure the crossing
correlation of the synchronized lasers with a homemade cross
correlator. After passing one laser beam through variable delay
line with a motor-driven roof reflector, both laser beams are
focused in a nonlinear crystal BBO to generate the sum-frequency
signal (SFG). Measuring the SFG signal while scanning the delay
line, we observe the cross-correlation curve. The FWHM of the
cross-correlation curve is about 90 fs. Subtracting the
contributions of pulse duration, we can deduce that the two lasers
are synchronized with a timing jitter less than 2 femtoseconds. We
also observe that the two lasers are able to keep on synchronizing
over 24 hours, which indicate that the laser system is stable for
our further implementation. The short pulse duration and little
timing jitter are sufficient to ensure the perfect interference of
two independent photons produced by synchronized laser pulses.

\begin{figure}[ptb]
\includegraphics[width=\columnwidth]{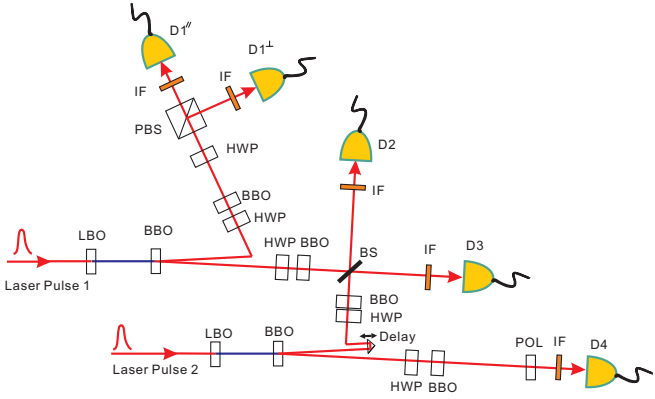}
\caption{The schematic drawing of experimental setup of quantum
entanglement swapping. Two 394 nm UV pulses are produced by
frequency doubling the 788 nm pulses of the synchronized lasers
using two nolinear LBO crystal ($LiB_3O_5$). Passing the UV pulses
through 2mm BBO ($\beta-BaB_2O_4$) crystals  creates two pairs of
polarization entangled photons in the entangled state
$|\Psi^-\rangle$, via type-II parametric down conversion. In order
to compensate the birefringence of the BBO crystals, we place a
half wave plate (HWP) and a compensating 1mm BBO crystal on each
path of the four photons. Interference filters (IF) with
$\Delta\lambda_{HMFW}$ = 2.8nm are place before each single photon
detector ($D_1-D_4$). The Beam splitter (BS) performs as a
Bell-state measurement here. To meet the condition of temporal
overlap, we used a step motor which minimum step is 0.1 $\mu m$ to
search for the position where the two photons arrived in the BS at
the same time. To verify this entanglement, we utilize a half wave
plate (HWP) and two detectors ($D_1^{\parallel}$ and
$D_1^{\perp}$) behind a polarizing beam splitter (PBS) to analyze
the polarization of photon 1. Photon 4 is analyzed by detector
$D_4$ behind a polarizer (POL) with a variable polarization
direction $\theta_4$.} \label{fig2}
\end{figure}

Figure \ref{fig2} is the schematic of the experimental setup of
entanglement swapping. Two 394 nm UV pulses are produced by
frequency doubling the 788 nm pulses of the synchronized lasers
using two nolinear LBO($LiB_3O_5$) crystal. For the first UV pulse
we obtained an average UV power of 250 mW, and for the second UV
pulse, 300mW. Passing the first UV pulses through a 2-mm-thick BBO
($\beta-BaB_2O_4$) crystal creates a pair of photons 1 and 2 in
the entangled state $|\Psi^{-}\rangle_{12}$, via type-II
parametric down conversion \cite{kwiat:4337}. For the 2-mm thick
BBO crystal, using filters with $\Delta\lambda_{HMFW}$ = 2.8nm,
the registered event rate of photon pairs was about 2000 count per
second. In the same way, another pair of photons 3 and 4 is
created by the second UV pulse in a different BBO crystal. For the
second pair of photons, again using filters with
$\Delta\lambda_{HMFW}$ = 2.8nm, we obtained 2500 count per second.
The observed visibility in the 45 degree polarization basis is
about $90\%$ for both photon pairs.

According to the entanglement swapping scheme, upon projection of
photons 2 and 3 into the $|\Psi^{-}\rangle_{23}$ state, photon 1
and 4 should be projected into $|\Psi^{-}\rangle_{14}$ state. To
verify that this entangled state is obtained, we have to analyze
the polarization correlation between photons 1 and 4 conditioned
on coincidences between the detectors ($D_2$ and $D_3$) of the
Bell-state analyzer. We utilize a half wave plate and two
detectors ($D_1^\parallel$ and $D_1^\perp$) behind a polarizing
beam splitter to analyze the polarization of photon 1. For
example, we can choose to analyze the polarization of photon 1
along the $+45^\circ$ and $-45^\circ$ by rotating the half wave
plate $22.5^\circ$. Photon 4 is analyzed by detector $D_4$ behind
a polarizer with a variable polarization direction $\theta_4$.

\begin{figure}[ptb]
\includegraphics[width=\columnwidth]{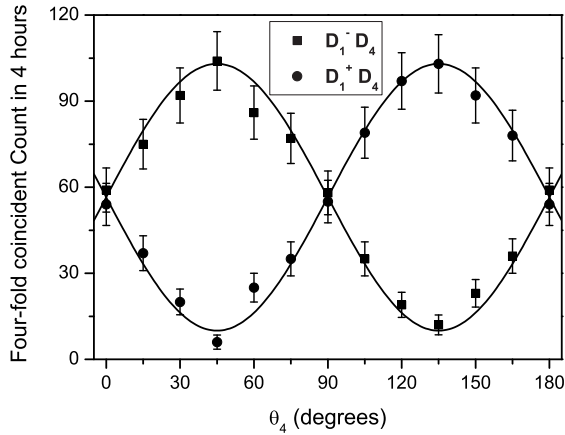}
\caption{Entanglement verification. Fourfold coincidences,
resulting from twofold $D_1^+$$D_4$ and $D_1^-$$D_4$ coincidences
conditioned on the twofold coincidences of the Bell-state
measurement. When varying the polarizer angle $\theta_4$, the two
complimentary sine curve with a visibility of 82\% demonstrate
that photons 1 and 4 are polarization entangled.} \label{fig3}
\end{figure}

If entanglement swapping happens, then the twofold coincident
between $D_1^\parallel$ and $D_4$, and $D_1^\perp$ and $D_4$,
conditioned on the $|\Psi^{-}\rangle_{23}$ detection, should show
two sine curves as a function of $\theta_4$ which are $90^\circ$
out of phase. Figure \ref{fig3} shows the experimental one of our
result for the coincidences between $D_1^\parallel$ and $D_4$, and
$D_1^\perp$ and $D_4$, given that photons 2 and 3 have been
registered by the two detectors in the Bell-state analyzer, where
we rotate the half wave plate $22.5^\circ$ to make $D_1^\parallel$
to register photon 1 with $+45^\circ$ polarization, and
$D_1^\perp$ to register photon 1 with $-45^\circ$ polarization.
The experimentally obtained four-fold coincidences shown in figure
\ref{fig3} have been fitted by a joint sine function with the same
amplitude for both curves. The observed visibility of $82\%$
clearly surpasses the 0.71 limit of Bell's inequalities, which
indicates the entanglement swapping do has been happened.

The high-visibility sinusoidal coincident curves in the experiment
imply a violation of a suitable  Bell's inequality. In particular,
according to the Clauser-Horne-Shimony-Holt (CHSH) inequality
\cite{Clauser:23880}, $S\leq2$ for any local realistic theory,
where
\begin{eqnarray}\label{eq:S}
S & = &
|E(\theta_1,\theta_4)-E(\theta_1,\theta_4')-E(\theta_1',\theta_4)-E(\theta_1',\theta_4')|,
\end{eqnarray}
and the $E(\theta_1,\theta_4)$ is the coefficient for measurement
where $\theta_1$ (or $\theta_1'$) is the polarizer setting for
photon 1, and $\theta_4$ (or $\theta_4'$) is the setting for
photon 4. In our experiment we set $\theta_1 = -22.5^\circ,
\theta_1' = -67.5^\circ, \theta_4 = 0^\circ, \theta_4' =
45^\circ$, which maximizes the quantum mechanics' prediction of
$S$ to $S^{QM} = 2\sqrt{2}$ and leads to a contradiction between
local realistic theory and quantum mechanics. In our experiment,
the four correlation coefficients between photons 1 and 4 gave the
follow results: $E(-22.5^\circ, 0^\circ) = -0.570 \pm 0.049,
E(-22.5^\circ, 45^\circ) = 0.583 \pm  0.046, E(-67.5^\circ,
0^\circ) = 0.600 \pm  0.049, E(-67.5^\circ, 45^\circ) = 0.554 \pm
0.046$. Hence, $S=2.308\pm 0.095$ which violates the classical
limit of 2 by 3.2 standard deviations. This clearly confirm the
quantum nature of entanglement swapping.

In summary, in the experiment we have exploited two synchronized
femtosecond lasers to report for the first time an experimental
demonstration of entanglement swapping with independent entangled
photon pairs. Whereas our experiment presents a strict
experimental realization of entangling photons that never
interacted, the techniques developed in the experiment can be
readily used to generate synchronized entangled photon pairs in
all segments by cascading the coupling between the lasers, hence
taking a significant step towards realistic linear optical
realization of quantum repeaters and quantum computation.

\begin{acknowledgments}
This work was supported by the National Natural Science Foundation
of China, Chinese Academy of Sciences and the National Fundamental
Research Program (under Grant No. 2001CB309300).
\end{acknowledgments}

\end{document}